\documentclass[twocolumn]{pasj00}
\SetRunningHead{Astronomical Society of Japan}{XMM-Newton and Suzaku
observations of X-ray stars towards the
Galactic bulge}

\usepackage{times}
\usepackage{ulem}
\usepackage{color}

\begin{document}

\title{XMM-Newton and Suzaku Spectroscopic Studies of Unidentified X-ray
Sources towards the Galactic Bulge: \\
1RXS J180556.1$-$343818 and 1RXS J173905.2$-$392615}
\author{Hideyuki \textsc{Mori},\altaffilmark{1,2}
Yoshitomo \textsc{Maeda},\altaffilmark{3}
Yoshihiro \textsc{Ueda},\altaffilmark{4}}
\altaffiltext{1}{CRESST and X-ray Astrophysics Laboratory, NASA Goddard
Space Flight Center, Greenbelt, MD 20771, USA}
\altaffiltext{2}{Department of Physics, University of Maryland,
Baltimore County, 1000 Hilltop Circle, Baltimore, MD 21250, USA}
\altaffiltext{3}{Department of Space Astronomy and Astrophysics,
Institute of Space and Astronautical Science (ISAS), Japan Aerospace
Exploration Agency (JAXA), 3-1-1, Yoshinodai, Chuo-ku, Sagamihara,
252-5210} \altaffiltext{4}{Department of Astronomy, Graduate School of
Science, Kyoto University, Sakyo-ku, Kyoto, 606-8502}
\email{hideyuki.mori@nasa.gov}

\maketitle

\begin{abstract}
With the XMM-Newton and Suzaku observations, for the first time, we
 acquired broad-band spectra of two unidentified X-ray sources towards
 the Galactic bulge: 1RXS J180556.1$-$343818 and 1RXS
 J173905.2$-$392615.  The 1RXS J180556.1$-$343818 spectrum in the
 $0.3$--$7$~keV band was explained by X-ray emission originated from an
 optically-thin thermal plasma with temperatures of $0.5$ and $1.7$~keV.
 The estimated absorption column density of $N_{\rm H} \sim 4 \times
 10^{20}$~cm$^{-2}$ was significantly smaller than the Galactic
 H\emissiontype{I} column density towards the source.  A candidate of
 its optical counterpart, HD 321269, was found within $\timeform{4''}$.
 In terms of the X-ray properties and the positional coincidence, it is
 quite conceivable that 1RXS J180556.1$-$343818 is an active G giant.
 We also found a dim X-ray source that was positionally consistent with
 1RXS J173905.2$-$392615.  Assuming that the X-ray spectrum can be
 reproduced with an absorbed optically-thin thermal plasma model with
 $kT = 1.6$~keV, the X-ray flux in the $0.5$--$8$~keV band was $8.7
 \times 10^{-14}$~erg~s$^{-1}$~cm$^{-2}$, fainter by a factor of $\sim
 7$ than that of 1RXS J173905.2$-$392615 during the ROSAT observation.
 The follow-up observations we conducted revealed that these two sources
 would belong to the Galactic disk, rather than the Galactic bulge.
\end{abstract}

\section{Introduction}
\label{section:intro}
A key to decipher the dynamical formation of the Galaxy is a population
study in the representative Galactic components: Galactic center, disk,
and the bulge.  The Galactic bulge is an old ($> 10$~Gyr) spheroidal
rotator with an angular size of $\sim 10^{\circ}$, where massive stars
are considered to have ended up to compact objects such as black holes
or neutron stars.  Since these objects produce enormous amounts of
X-rays ($L_{\rm X} > 10^{32}$~erg~s$^{-1}$) via the accretion process,
hence, the luminosity function of luminous X-ray sources provides
indirect information on the distribution of primordial high-mass stars
in the Galactic bulge.  Because of its proximity, deep investigations in
some intriguing regions were devoted to revolve spatially X-ray sources
down to the X-ray fluxes of several times
$10^{-14}$~erg~s$^{-1}$~cm$^{-2}$ (e.g., Galactic Bulge Survey;
\cite{2011ApJS..194...18J}), which have never been achieved in the
extragalactic survey\footnote{The X-ray flux of
$10^{-14}$~erg~s$^{-1}$~cm$^{-2}$ corresponds to the X-ray luminosity of
$\sim 10^{32}$~erg~s$^{-1}$ at a distance of $8.5$~kpc, which is $\sim
2$ orders of magnitude smaller than the sensitivity limit of the Chandra
survey of the M31 bulge ($4 \times 10^{34}$~erg~s$^{-1}$;
\cite{2011A&A...533A..33Z})}.

However, once we focus on moderately bright X-ray sources in the
Galactic bulge, which fluxes of $\sim 10^{-12}$~erg~s$^{-1}$~cm$^{-2}$
correspond to the X-ray luminosity of $L_{\rm X} \sim
10^{34}$~erg~s$^{-1}$ at $8.5$~kpc, the low space density of these
sources requires comprehensive mapping observations with wide-field
coverage and relatively long exposures.  This motivates us to develop an
approach to pick up the discrete X-ray sources in the Galactic bulge
from the ROSAT All-Sky Survey (RASS) data.  Using the X-ray color
provided in the RASS Bright Source Catalogue (RBSC;
\cite{1999A&A...349..389V}), we made a flux-limited sample of 68 sources
towards the Galactic bulge; the nature of about half of these RBSC
sources remains unclear.  In order to improve the completeness of the
Galactic bulge X-ray sources and to construct the luminosity functions
of different populations in the Galactic bulge, we now proceed a
campaign of X-ray follow-up observations for spectroscopic
identification to the unidentified X-ray sources.  A total of 21 RBSC
sources has been observed so far with Suzaku (5), XMM-Newton (5), and
Chandra (11).  On the basis of these follow-up observations, 3 low-mass
X-ray binaries (\cite{2005A&A...440..287I}, \cite{2009A&A...502..905Y},
\cite{2012PASJ...64..112M}), 1 RS CVn star \citep{2012PASJ...64..112M},
and 1 cluster of galaxies \citep{2013PASJ...65..102M} were identified.

In this paper, we report the results of two unidentified X-ray sources
obtained with the XMM-Newton and Suzaku observations; these two sources
turned out most likely to be foreground stars.  Although these sources
are unlikely to be the Galactic bulge population, these stars are
available for understanding the X-ray population in the Galactic disk;
our result may provide a clue to resolve the enhanced soft X-ray
emission below $1$~keV, so-called ``M-band problem'' (e.g.,
\cite{1990ARA&A..28..657M}), or the Galactic Ridge X-ray Emission (GRXE;
\cite{1986PASJ...38..121K}).

We describe the observation and data reduction in
section~\ref{section:observation}.  The analysis of the X-ray images and
spectra as well as their results are given in
section~\ref{section:analysis}.  After the discussion on the nature of
the X-ray sources in section~\ref{section:discussion}, we summarize our
results in section~\ref{section:summary}.  We note that errors represent
the $90$\% confidence limits throughout the paper, unless otherwise
mentioned.

\section{Observation and data reduction}
\label{section:observation}
Both 1RXS J180556.1$-$343818 and 1RXS J173905.2$-$392615 were first
discovered with the RASS.  The positions of these sources in Galactic
coordinates are $(l, b) = (\timeform{357.219D}, \timeform{-6.589D})$ for
1RXS J180556.1$-$343818 and $(\timeform{350.349D}, \timeform{-4.373D})$
for 1RXS J173905.2$-$392615.  The source count rates in the ROSAT PSPC
band of 1RXS J180556.1$-$343818 and 1RXS J173905.2$-$392615 were $0.29
\pm 0.05$~cts~s$^{-1}$ and $0.07 \pm 0.02$~cts~s$^{-1}$, respectively.
In order to estimate X-ray fluxes in the $0.5$--$10$~keV band, we
assumed an absorbed power-law model with photon index of $\Gamma = 1.7$.
From the Galactic latitude ($b$) and the X-ray color for each source,
the absorption column density can be estimated to $N_{\rm H} =
(1.5$--$7.9) \times 10^{21}$~cm$^{-2}$ for 1RXS J180556.1$-$343818 and
$N_{\rm H} = (3.6$--$11.7) \times 10^{21}$~cm$^{-2}$ for 1RXS
J173905.2$-$392615.  Thus, the count rates for 1RXS J180556.1$-$341838
and 1RXS J173905.2$-$392615 correspond to the absorption-corrected X-ray
fluxes in the $0.5$--$10$~keV band of $(0.9$--$2.7) \times 10^{-11}$ and
$(0.3$--$1.0) \times 10^{-11}$~erg~s$^{-1}$~cm$^{-2}$, respectively.

In order to obtain the broad-band X-ray spectrum, we performed the
observation of 1RXS J180556.1$-$343818 with XMM-Newton
\citep{2001A&A...365L...1J} in March 2004.  XMM-Newton possesses
European Photon Imaging Camera (EPIC) placed on the focus of the X-ray
telescopes \citep{2000SPIE.4012..731A} with large photon collecting
area.  Two of the EPIC cameras consist of 7 CCD arrays manufactured with
Metal Oxide Semiconductor (MOS; \cite{2001A&A...365L..27T}).  Another
EPIC camera has an array of the pn-type CCD chips (pn;
\cite{2001A&A...365L..18S}).  The observation log is given in
table~\ref{table:observation}.  All of the EPIC cameras were operated in
the full frame mode with medium filters.  The exposure times during the
observation were $10.0$~ks for the pn camera and $11.6$~ks for the MOS
cameras.  The Science Analysis Software (version 13.0.1) was used to
reduce the data and to create scientific products.  We first reprocessed
the Observation Data Files, provided by the XMM-Newton Science Archive,
by running \texttt{epchain} and \texttt{emchain} tasks with the latest
calibration database.

To remove the background flares produced by soft protons, we selected
good time intervals (GTIs) by following the instruction in the XMM data
reduction
guide\footnote{http://xmm.esac.esa.int/external/xmm\_user\_support/documentation/sas\_usg/USG.pdf}.
We filtered out the time intervals in which the count rate of single
events (\texttt{PATTERN} = 0) over the full detector field of view was
larger than $0.4$~cts~s$^{-1}$ in the $10$--$12$~keV band for pn or
$0.35$~cts~s$^{-1}$ above $10$~keV for MOS1/2.  The flare-screened live
times for the pn, MOS1, and MOS2 cameras were $384$~s, $2.84$~ks, and
$2.94$~ks, respectively.

1RXS J173905.2$-$392615 was observed with the Suzaku satellite
\citep{2007PASJ...59S...1M} in September 2012.  The X-ray CCD cameras
(XIS; \cite{2007PASJ...59S..23K}) and the corresponding X-ray telescopes
(XRT; \cite{2007PASJ...59S...9S}) onboard Suzaku enable us to perform
imaging spectroscopy in the $0.2$--$10$~keV band.  The XIS consists of
one back-illuminated (BI) CCD and three front-illuminated (FI) ones.
However, one FI CCD (XIS2) has been out of order since November 9, 2006.
Furthermore, a part of the imaging area of the XIS0 has been unusable
since June 23, 2009\footnote{Suzaku memo: JX-ISAS-SUZAKU-MEMO-2010-01}.
The XIS was operated in the normal mode without any window and timing
options.  During the observation, the spaced-row charge injection
\citep{2009PASJ...61S...9U} was applied.  Suzaku also equips non-imaging
Hard X-ray Detector (HXD; \cite{2007PASJ...59S..35T},
\cite{2007PASJ...59S..53K}), consisting of the Si-PIN photodiodes and
GSO scintillators to cover the energy band of $10$--$600$~keV.  The HXD
was also operated in the normal mode.  The log of the Suzaku observation
is also given in table~\ref{table:observation}.

We reduced the cleaned event data which were pre-processed with the
processing version of 2.8.16.34.  Because of high Non-X-ray Backgrounds
(NXBs), time intervals during the passage of the South Atlantic Anomaly,
the low-Earth orbit, and the low Day-Earth orbit were discarded in the
cleaned event data.  The effective exposure time was $22.1$~ks for the
XIS and $17.1$~ks for the HXD.  The reduction tool we used here was
HEASOFT (version 6.13).  X-ray images, light curves, and spectra were
generated by running \texttt{Xselect}.  We note that the HXD-PIN count
rate in the $18$--$40$~keV band was $0.157 \pm 0.003$~cts~s$^{-1}$. It
was consistent with that of the NXB ($0.155 \pm 0.001$~cts~s$^{-1}$)
derived from the ``tuned background'' file,
\texttt{ae407029010\_hxd\_pinbgd.evt.gz}, provided from the HXD team.
Since significant signals were not detected from the HXD-PIN, we focused
on the XIS data in the following analysis.

\begin{table*}[bhtp]
\begin{center}
\caption{Observation log.}
\label{table:observation}
\begin{tabular}{llccc}
\hline
Target & Observatory & Obs. ID & Start / End time (UT) &
 Exposure (ks)\footnotemark[$*$] \\
\hline
1RXS J180556.1$-$343818 & XMM-Newton & 0206990301 & 2004/03/11 07:44:37 & $2.8$ (MOS1), $2.9$ (MOS2)  \\
 & & & 2004/03/11 10:31:53 & $0.38$ (pn) \\
1RXS J173905.2$-$392615 & Suzaku & 407029010 & 2012/09/18 19:46:11 & $22.1$ (XIS), $17.1$ (PIN) \\
 & & & 2012/09/19 08:16:22 & \\
\hline
\multicolumn{5}{@{}l@{}}{\hbox to 0pt{\parbox{180mm} {\footnotesize
 \footnotemark[$*$] Effective exposure of the screened data. The HXD-PIN exposure is dead-time corrected.
 }\hss}}
\end{tabular}
\end{center}
\end{table*}

\section{Analysis}
\label{section:analysis}
\subsection{1RXS J180556.1$-$343818}
\label{subsection:1RXSJ180556}

\begin{figure*}[htbp]
 \begin{center}
  \FigureFile(80mm,50mm){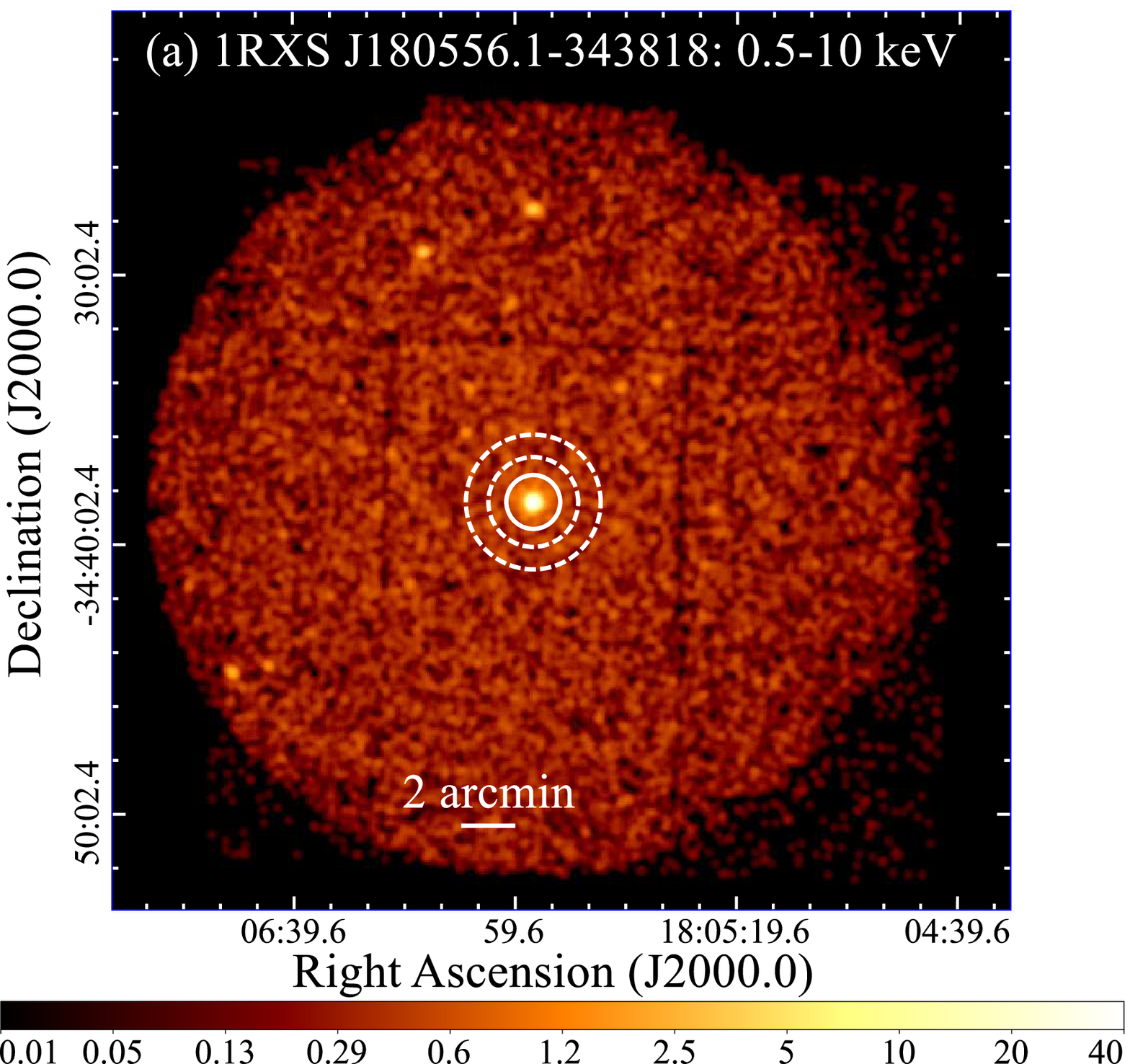}
  \FigureFile(80mm,50mm){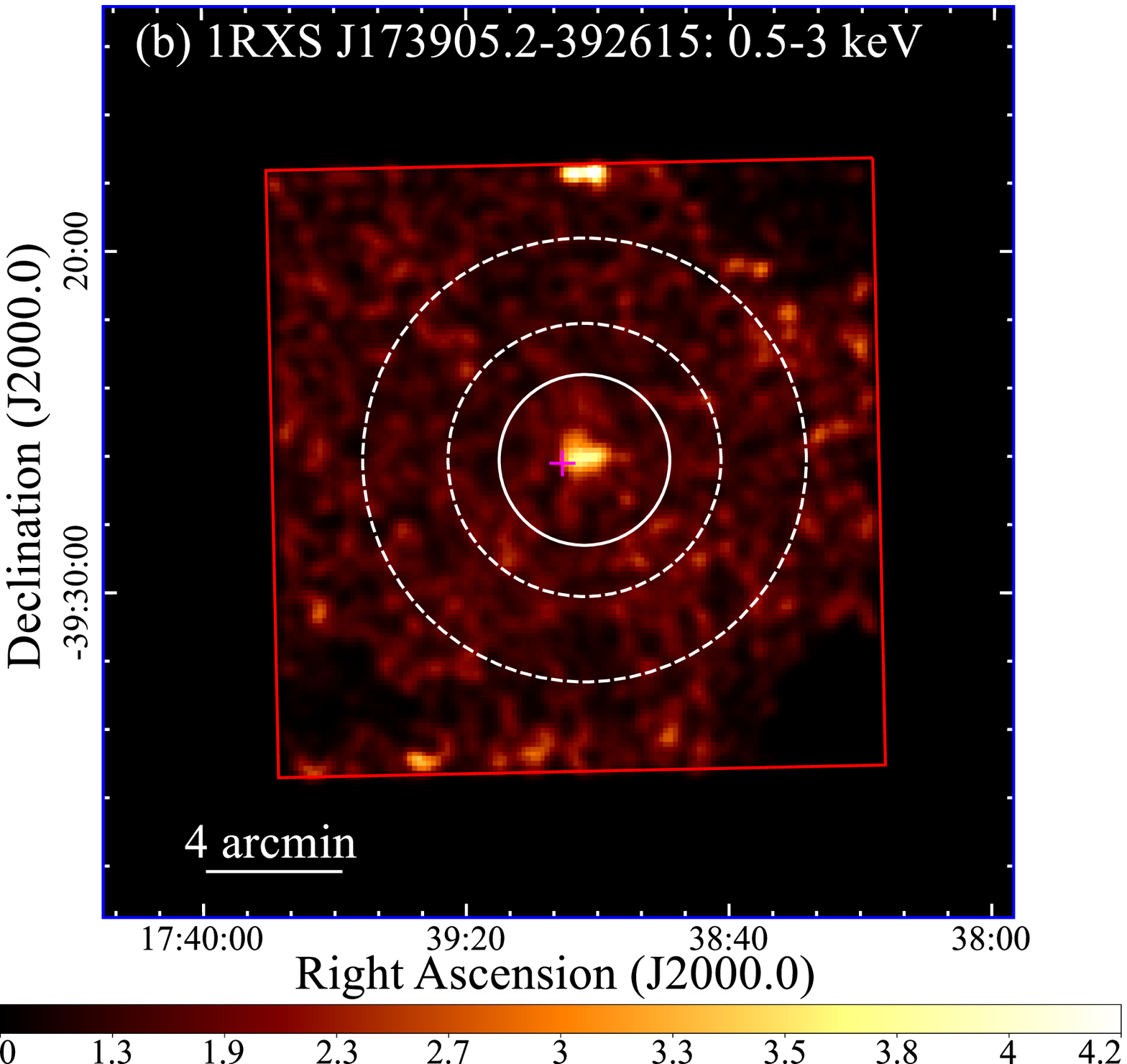}
 \end{center}
 \caption{
X-ray images of 1RXS J180556.1$-$343818 (a, $0.5$--$10$~keV) and 1RXS
 J173905.2$-$392615 (b, $0.5$--$3$~keV) obtained with the XMM-Newton and
 Suzaku observations.  The MOS1, MOS2, and pn images were binned with
 $\timeform{5''} \times \timeform{5''}$ and then added together.  The
 smoothing with the Gaussian function of $\sigma = \timeform{15''}$ was
 applied to the summed image.  No vignetting correction was performed.
 The solid white circle and the dashed white annulus centered on 1RXS
 J180556.1$-$343818 indicate the regions to extract the source and
 background spectra, respectively.  The XIS image was also binned with
 $8 \times 8$~pixels and then smoothed with the Gaussian kernel of
 $\sigma = \timeform{33''}$. The subtraction of the Non-X-ray Background
 and the vignetting correction were applied to the XIS image.  The red
 square represents the XIS field of view.  The source and background
 regions to extract the XIS spectra are indicated with the solid white
 circle and the dashed white annulus, respectively.  The position of the
 RBSC source is shown with the magenta plus located in the eastern
 direction of the X-ray emission.  The color scales at the bottom panels
 are in unit of photons per $\timeform{5''} \times \timeform{5''}$ (a)
 and $\timeform{8.3''} \times \timeform{8.3''}$ (b) image pixel.
}
\label{fig:Xray_images}
\end{figure*}

Figure~\ref{fig:Xray_images}a shows the X-ray image in the
$0.5$--$10$~keV band obtained with the XMM-Newton observation.  We
created the MOS1, MOS2, and pn images with a pixel size of
$\timeform{5''} \times \timeform{5''}$ from the respective event files.
These images were added together, and then smoothed with the Gaussian
function of $\sigma = \timeform{15''}$.  1RXS J180556.1$-$343818 was
found at the center of the MOS 1/2 field of view.  Some sources outside
the error radius of 1RXS J180556.1$-$343818 were also detected, but
their detailed spectral analysis is beyond the scope of this paper.  The
position of the brightness peak was at $(RA, Dec)_{\rm J2000.0} =
(\timeform{18h05m56.33s}, \timeform{-34D38'29.3''})$.  Hence, as a
source-extraction region, we selected a circle with a radius of
$\timeform{1'}$ centered on this position.  We chose the surrounding
source-free annulus as a background region. The inner and outer radii of
the annulus were set to be $\timeform{1.67'}$ and $\timeform{2.5'}$,
respectively.

For each sensor, we made the X-ray light curves in the $0.5$--$10$~keV
band from the source and background regions.  No significant variability
was found in the background-subtracted light curve.  Hence, we did not
apply any additional time filters in the spectral analysis.  By fitting
a constant model, we estimated the average count rates for MOS1, MOS2,
and pn to be $0.11 \pm 0.01$, $0.11 \pm 0.01$, and $0.30 \pm
0.04$~cts~s$^{-1}$, respectively.

\begin{figure}[htbp]
 \begin{center}
   \FigureFile(80mm,50mm){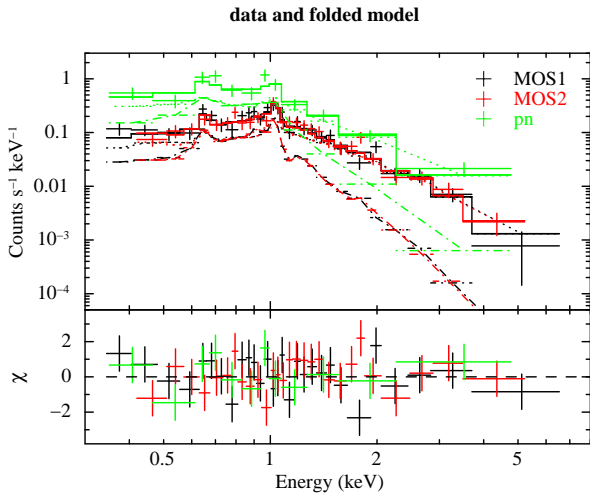}
 \end{center}
 \caption{X-ray spectra of 1RXS J180556.1$-$343818. The spectra obtained
 with MOS1, MOS2, and pn are shown with black, red, and green pluses.
 Solid lines overlaid on the respective spectra represent the best-fit
 model consisting of the two-temperature CIE plasma components.  The
 low-temperature and high-temperature components are indicated with
 dash-dotted and dotted lines, respectively.}
 \label{fig:1RXSJ180556_spectra}
\end{figure}

The X-ray spectra were extracted from the source and background regions
with \texttt{evselect}.  The redistribution matrix file (RMF) and
ancillary response file (ARF) were created with \texttt{rmfgen} and
\texttt{arfgen}, respectively.  We binned these spectra so that each bin
contains at least $20$~photons.  The background-subtracted X-ray spectra
are shown in figure~\ref{fig:1RXSJ180556_spectra}.  Since the
significant photons were found at $< 7$~keV, we modelled the spectra in
the $0.3$--$7$~keV band.

First, we tried to fit the X-ray spectra with a power-law or blackbody
model attenuated with the photoelectric absorption
\citep{1983ApJ...270..119M}.  The abundance pattern we selected was that
derived from \citet{1989GeCoA..53..197A}.  We multiplied the model by a
constant, which is fixed at unity for the FI CCDs but is left free for
the BI CCD, to absorb the uncertainty of the absolute flux calibration
in each instrument.  Both simple models were unacceptable; the power-law
model with $\Gamma = 3.5$ and the blackbody model with $kT = 0.3$~keV
gave the reduced $\chi^{2}$ of $2.2$ and $2.7$, respectively.  We found
a clear residual around $1$~keV, indicating the presence of the
K$\alpha$ emission line from highly ionized Ne.  This suggests that an
optically-thin thermal plasma was responsible for the X-ray emission.

Therefore, we next tried to fit the spectra with the \texttt{apec} model
that reproduces X-ray emission originated from an optically-thin thermal
plasma in collisional ionization equilibrium (CIE).  Although the fit
was improved, it was still unacceptable in terms of the reduced $\chi
^{2}$ ($103/63$~d.o.f. = $1.6$).  Moreover, the best-fit metal abundance
of $Z/Z_{\odot} = 0.1$ did not fully reproduce the Ne\emissiontype{X}
emission line and the structure around $0.6$--$0.7$~keV.  Then, we
replaced the \texttt{apec} model into the \texttt{vapec} model to
determine each elemental abundance individually.  Here, O, Ne, Si, S,
and Fe abundances were thawed; the other abundances were fixed at $0.1$
times the solar values.  This model yielded an acceptable fit of $\chi
^{2} = 65$ ($59$~d.o.f.).  The best-fit parameters of the plasma
temperature and the absorption column density were $kT = 1.2$~keV and
$N_{\rm H} = 4.0 \times 10^{20}$~cm$^{-2}$, respectively.  O, Ne, and S
were close or overabundant to the solar values ($Z/Z_{\odot} =
0.9$--$1.8$), whereas the Si and Fe abundances did not change
significantly from $Z/Z_{\odot} = 0.1$.

To investigate whether the enhanced abundances of O, Ne and S were an
artifact due to the single-temperature plasma model, we then added
another thermal plasma component to the model.  The absorption column
density and the metal abundance were tied between the two plasma
components.  While we again allowed the elemental abundances for O, Ne,
Mg, Si, and Fe to vary, the other abundances were set to $Z/Z_{\odot} =
0.1$.  The two-temperature \texttt{vapec} model also gave an acceptable
fit with $\chi ^{2} = 56$ ($57$~d.o.f.).  The best-fit model was
superposed on the X-ray spectra in figure~\ref{fig:1RXSJ180556_spectra}.
Since the $F$-test yielded the chance probability of $P = 1$\% due to
statistical fluctuations, this two-temperature plasma model was found to
improve the fit significantly.  The absorption column density of $N_{\rm
H} = 3.3 \times 10^{20}$~cm$^{-2}$ was consistent with that derived from
the single-temperature thermal plasma model within the error.  The
best-fit parameters for the plasma temperatures were $kT =
0.5^{+0.5}_{-0.1}$ and $1.8^{+0.5}_{-0.4}$~keV.  The O, Ne, Si, S and Fe
abundances relative to solar values were $1.1$, $2.2$, $0.3$, $1.0$, and
$0.08$, respectively.  We note that, when the metal abundance except for
O, Ne, Si, S, and Fe was changed to be $Z/Z_{\odot} = 0.3$, the best-fit
parameters for $N_{\rm H}$, $kT$, and the elemental abundances were
consistent with those derived from the case of $Z/Z_{\odot} = 0.1$.  In
the case of setting all of the abundances to be free, the trend of
sub-solar Fe abundance did not change, though meaningful constraints
cannot be given to the metal abundances.

We further checked the abundances of O, Ne, and S using a
multi-temperature plasma model (\texttt{cevmkl} in \texttt{XSPEC}).  In
this model, emission measures (EMs) are given as a function of plasma
temperatures: $d{\rm EM}/dT \propto (1/T_{\rm max})(T/T_{\rm
max})^{\alpha - 1} $.  Here we can specify the maximum temperature
($T_{\rm max}$) and the power-law index ($\alpha$).  This
multi-temperature plasma model was also statistically acceptable ($\chi
^{2} = 65$ ($58$~d.o.f.)).  The best-fit maximum temperature and the
power-law index were $kT_{\rm max} = 2.0$~keV and $\alpha = 1.8$,
respectively.  Although the elemental abundances for Si and Fe were
subsolar values ($Z/Z_{\odot} \sim 0.1$), the O, Ne, and S abundances
were found to be $0.9$, $1.6$, and $0.8$ times the solar values,
respectively.  We note that the Fe abundance was in the range of
$0.02$--$0.34$ when we re-fitted the spectra using the abundance table
of \citet{2003ApJ...591.1220L}, instead of that of
\citet{1989GeCoA..53..197A}.  Furthermore, the Fe abundance was
estimated to be $Z/Z_{\odot} = 0.3^{+0.7}_{-0.2}$ when all of the
abundances were set to be free.

The best-fit parameters obtained with the two-temperature and
multi-temperature plasma models were summarized in
table~\ref{table:1RXSJ180556_fit}.  The X-ray flux in the $0.3$--$7$~keV
band was $1.4 \times 10^{-12}$~erg~s$^{-1}$~cm$^{-2}$.

\begin{table}
\begin{center}
\caption{Best-fit parameters of the 1RXS J180556.1$-$343818 spectra\footnotemark[$*$].}
\label{table:1RXSJ180556_fit}
\begin{tabular}{lcc}
\hline
Parameters  & \multicolumn{2}{c}{Model\footnotemark[$\dagger$]} \\
            & \texttt{vapec}+\texttt{vapec} & \texttt{cevmkl} \\
\hline
$N_{\rm H}$ ($10^{20}$ cm$^{-2}$) & $3.3 (< 9.1)$ & $4.1 (< 9.6)$ \\
$kT_{\rm 1}$ (keV)                & $0.5^{+0.5}_{-0.1}$ & --- \\
$kT_{\rm 2}$ (keV)                & $1.8^{+0.5}_{-0.4}$    & --- \\
$kT_{\rm max}$ (keV)              & --- & $2.0^{+1.0}_{-0.6}$ \\
$\alpha$\footnotemark[$\ddagger$] & --- & $1.8^{+1.6}_{-0.8}$ \\
$Z_{\rm O}$\footnotemark[$\S$]    & $1.1^{+2.1}_{-0.7}$ & $0.9^{+0.9}_{-0.5}$ \\
$Z_{\rm Ne}$\footnotemark[$\S$]   & $2.2^{+3.6}_{-1.5}$ & $1.6^{+1.6}_{-0.9}$ \\
$Z_{\rm Si}$\footnotemark[$\S$]    & $0.3 (< 1.1)$ & $0.2 (< 0.8)$ \\
$Z_{\rm S}$\footnotemark[$\S$]    & $1.0^{+1.5}_{-0.9}$ & $0.8 (< 1.9)$ \\
$Z_{\rm Fe}$\footnotemark[$\S$]   & $0.08^{+0.21}_{-0.07}$ & $0.14^{+0.21}_{-0.10}$ \\
${\rm Normalization}_{\rm 1}$\footnotemark[$\l$]   & $\bigl ( 4^{+15}_{-3} \bigr )
     \times 10^{-4}$ & $\bigl ( 6.5^{+5.2}_{-2.8} \bigr ) \times 10^{-3}$  \\
${\rm Normalization}_{\rm 2}$\footnotemark[$\l$]   & $\bigl ( 1.3^{+0.4}_{-0.5} \bigr ) \times 10^{-3}$ & --- \\
\hline
$\chi ^{2}$/d.o.f.                & $56/57 = 0.98$ & $65/58 = 1.1$\\
Flux ($0.3$--$7$~keV)       & \multicolumn{2}{c}{$1.4 \times 10^{-12}$~erg s$^{-1}$ cm$^{-2}$} \\
\hline
\multicolumn{3}{@{}l@{}}{\hbox to 0pt{\parbox{80mm} {\footnotesize
 \footnotemark[$*$] Superscript and subscript figures represent the
upper and lower limits of the 90\% confidence interval, respectively.
Figures in parentheses indicate the 90\% upper limit.
 \par\noindent
 \footnotemark[$\dagger$] Model components defined in \texttt{XSPEC}.
 \par\noindent
 \footnotemark[$\ddagger$] Power-law index for calculating the
differential emission measure at temperature $T$.
 \par\noindent
 \footnotemark[$\S$] Elemental abundances relative to solar \citep{1989GeCoA..53..197A}.
 \par\noindent
 \footnotemark[$\l$] In unit of $10^{-14}/(4 \pi D^{2}) \int n_{\rm e}
n_{\rm H} dV$~cm$^{-5}$. Here $V$ and $D$ are the volume and distance
to the plasma, respectively.
}\hss}}
\end{tabular}
\end{center}
\end{table}

\subsection{1RXS J173905.2$-$392615}
\label{subsection:1RXSJ173905}
The XIS image of the 1RXS J173905.2$-$392615 field in the
$0.5$--$3.0$~keV band is shown in figure \ref{fig:Xray_images}b.  We
made an XIS image with a binning of $8 \times 8$ pixels for each sensor.
After the corresponding NXB image created with \texttt{xisnxbgen}
\citep{2008PASJ...60S..11T} was subtracted, we divided the image by a
sensitivity map at $1$~keV to mitigate the XRT's vignetting effect.  The
sensitivity map was produced with \texttt{xissim}
\citep{2007PASJ...59S.113I}.  The vignetting-corrected images were then
added together.  Finally, the summed image was smoothed with the
Gaussian function with $\sigma = \timeform{33''}$.  X-ray emission was
detected near the position of the RBSC source (magenta plus in
figure~\ref{fig:Xray_images}b).  The peak of the X-ray brightness was
located at $(RA, Dec)_{\rm J2000.0} = (\timeform{17h39m02.0s},
\timeform{-39D26'07''})$.

\begin{figure}[htbp]
 \begin{center}
   \FigureFile(80mm,50mm){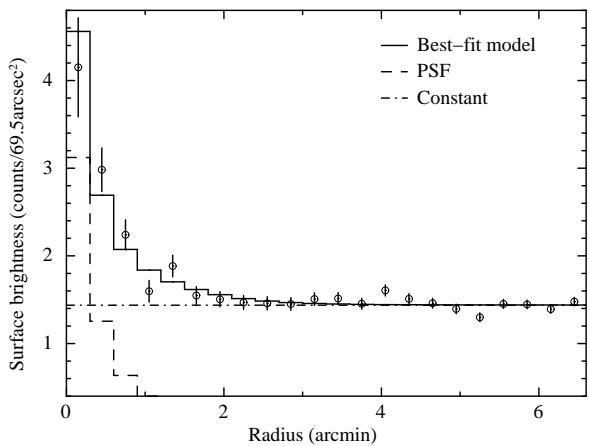}
 \end{center}
 \caption{Radial profile of the surface brightness of Suzaku
J1739$-$3926.  The vertical bars on the data represent a $1 \sigma$
statistical error.  The solid line indicates the best-fit model
consisting of the Point Spread Function of the XRT (dashed line) and a
constant (dash-dotted line).}
 \label{fig:1RXSJ173905_radial_profile}
\end{figure}

The X-ray source, designated as Suzaku J1739$-$3926 hereafter, was
extremely dim.  Thus, we made a radial profile of the surface brightness
(see figure~\ref{fig:1RXSJ173905_radial_profile}) from the XIS image in
the $0.5$--$3$~keV band.  Here we set the origin of the radial profile
to the position of the peak brightness.  We fitted the radial profile
within $\timeform{7'}$ with the Point Spread Function (PSF) of the XRT
plus a constant model.  The PSF was derived from the XIS image of SS Cyg
that was used for the in-flight calibration \citep{2007PASJ...59S...9S}.
The peak of the PSF was normalized to be unity.  We then set the PSF
normalization and the constant to be free parameters.  This model gave a
marginally acceptable fit of $\chi ^{2} = 29$ ($20$~d.o.f.).  The
best-fit parameters were $3.1$ for the PSF normalization and $1.4$ for
the constant.  Due to the limited photon statistics, we cannot ruled out
the possibility that multiple discrete sources or diffuse hot gas
contribute to Suzaku J1739$-$3926.  However, in the following analysis,
we presumed that the emission comes from a single point source.

In order to determine the source position more accurately, we fitted a
2-dimensional PSF model to the vignetting-corrected image before
applying the Gaussian smoothing.  The PSF model was constructed from an
analytical form that consists of the 3 Gaussian and 1 exponential
components, which well reproduced the PSF (dashed line in
figure~\ref{fig:1RXSJ173905_radial_profile}) within $\timeform{4'}$.  A
$2 \times 2$ binning was applied to the image to increase photon
statistics per each pixel for the $\chi ^{2}$ minimization.  We set the
normalization and the center of the PSF to be free parameters.  The
best-fit position of Suzaku J1739$-$3926 was $(RA, Dec)_{\rm J2000.0} =
(\timeform{17h39m02.8s}, \timeform{-39D25'56''})$.  The PSF fit gave the
90\% errors in the $RA$ and $Dec$ directions as $\timeform{5''}$ (east),
$\timeform{12''}$ (west), $\timeform{4''}$ (north), and $\timeform{7''}$
(south).  In addition, taking into account the absolute positional
uncertainty for Suzaku pointing observations
\citep{2008PASJ...60S..35U}, we estimated the positional accuracy of
Suzaku J1739$-$3926 to be $\sim \timeform{22''}$.

The source and background regions to extract the XIS spectra are shown
by a solid circle and a dashed annulus in figure~\ref{fig:Xray_images}b,
respectively.  On the basis of the radial profile of the surface
brightness, we determined the radius of the source-extraction region to
be $\timeform{2.5'}$.  The inner and outer radii of the annulus for the
background region were set to be $\timeform{4'}$ and $\timeform{6.5'}$,
respectively.  The RMF and ARF were created with \texttt{xisrmfgen} and
\texttt{xissimarfgen} \citep{2007PASJ...59S.113I}, respectively.  Suzaku
J1739$-$3926 was too faint to obtain accurate source spectra by
subtracting the background ones applied to only the area correction; the
contribution of celestial X-ray backgrounds in the source-extraction
region was underestimated because of the telescope vignetting.  Since
the source is located at low Galactic latitude of $b =
\timeform{-4.373D}$, the X-ray spectra are contaminated from not only
the Cosmic X-ray Background (CXB) but also the GRXE.  The GRXE is
characteristic of intense K$\alpha$ emission lines from highly ionized
Fe ions.  Thus, the underestimated background spectra would leave a
false structure around $7$~keV in the source spectra.

\begin{figure}[htbp]
 \begin{center}
   \FigureFile(80mm,50mm){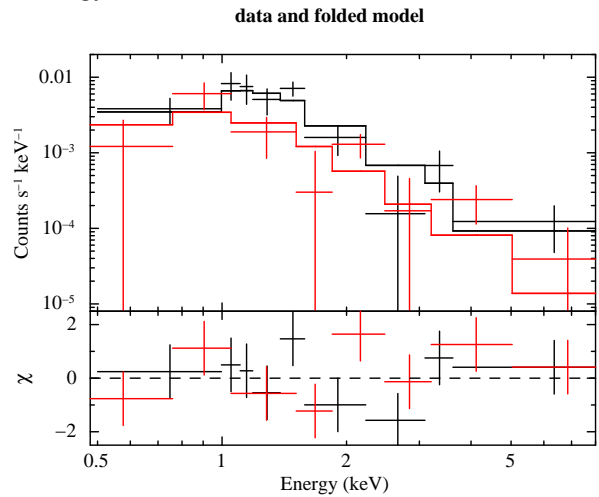}
 \end{center}
 \caption{Background-subtracted XIS spectra of Suzaku J1739$-$3926; FI
 and BI spectra are shown with black and red pluses, respectively.  The
 solid lines represent the best-fit power-law model.}
 \label{fig:1RXSJ173905_spectrum}
\end{figure}

We then generated new ARFs for the source and background regions,
assuming a uniform sky with a radius of $\timeform{20'}$, designated as
$A_{\rm src}$ and $A_{\rm bgd}$, respectively.  The NXB spectra for the
source ($N_{\rm src}$) and background ($N_{\rm bgd}$) regions were
created with \texttt{xisnxbgen}.  Next, for the area correction
including the vignetting effect, we made background-subtracted spectra
as follows; the count rate in each energy bin, $S$, was given by
\begin{equation}
S = \bigl ( S_{\rm src} - N_{\rm src} \bigr ) - \frac{A_{\rm
src}}{A_{\rm bgd}} \times \bigl ( S_{\rm bgd} - N_{\rm bgd} \bigr ),
\end{equation}
where $S_{\rm src}$, $S_{\rm bgd}$ represent the raw count rates
extracted from the source and background regions, respectively.  The
XIS0 and XIS3 spectra were added together to improve the photon
statistics.  We show the background-subtracted spectra in
figure~\ref{fig:1RXSJ173905_spectrum}.  These spectra were binned to
contain at least $10$~photons per each energy bin.

We fitted the spectrum in the $0.5$--$8$~keV band with a power-law model
(\texttt{powerlaw} in \texttt{XSPEC}) with photoelectric absorption.
This model gave an acceptable fit with $\chi ^{2} = 15$ ($13$~d.o.f.).
The best-fit parameters were $N_{\rm H} = 1.3 \times 10^{21}$~cm$^{-2}$
and $\Gamma = 3.0$.  However, because of poor photon statistics, the
absorption column density cannot be constrained significantly; only the
upper limit of $N_{\rm H} < 4.8 \times 10^{21}$~cm$^{-2}$ was obtained.
The photon index indicated the steep soft X-ray spectra.  Thus, we also
employed a CIE plasma model (\texttt{apec} in \texttt{XSPEC}) with
photoelectric absorption.  The metal abundance was fixed to be
$Z/Z_{\odot} = 0.3$.  An acceptable fit was yielded again; while the
best-fit temperature was found to be $kT = 1.6^{+1.0}_{-0.9}$~keV, the
upper limit of $N_{\rm H} < 6.8 \times 10^{21}$~cm$^{-2}$ was derived
for the absorption column density.  The best-fit parameters were
summarized in table~\ref{table:1RXSJ173905_fit}.  In order to estimate
the correct X-ray flux, we took into account a fraction of the photon
flux within the source-extraction region using the XRT's encircled
energy function \citep{2007PASJ...59S...9S}.  The X-ray flux in the
$0.5$--$8$~keV band was $8.7 \times 10^{-14}$~erg~s$^{-1}$~cm$^{-2}$.

\begin{table}
\begin{center}
\caption{Best-fit parameters of the Suzaku J1739$-$3926 spectra\footnotemark[$*$].}
\label{table:1RXSJ173905_fit}
\begin{tabular}{lcc}
\hline
Parameters  & \multicolumn{2}{c}{Model\footnotemark[$\dagger$]} \\
            & \texttt{powerlaw} & \texttt{apec} \\
\hline
$N_{\rm H}$ ($10^{21}$ cm$^{-2}$) & $1.3 (< 4.8)$ & $0 (< 6.8)$ \\
$\Gamma$\footnotemark[$\ddagger$] & $3.0^{+1.6}_{-0.7}$ & --- \\
$kT$ (keV)                & --- & $1.6^{+1.0}_{-0.9}$ \\
$Z/Z_{\odot}$\footnotemark[$\S$] & --- & $0.3$ (fixed) \\
\hline
$\chi ^{2}$/d.o.f.                & 15/13 $=$ 1.2 & 16/13 $=$ 1.3\\
Flux ($0.5$--$8$~keV)       & \multicolumn{2}{c}{$8.7 \times 10^{-14}$~erg s$^{-1}$ cm$^{-2}$} \\
\hline
\multicolumn{3}{@{}l@{}}{\hbox to 0pt{\parbox{70mm} {\footnotesize
 \footnotemark[$*$] Superscript and subscript figures represent the
upper and lower limits of the 90\% confidence interval, respectively.
 \par\noindent
\footnotemark[$\dagger$] Model components defined in \texttt{XSPEC}.
 \par\noindent
 \footnotemark[$\ddagger$] Photon index of the power-law continuum.
 \par\noindent
 \footnotemark[$\S$] Elemental abundances relative to solar \citep{1989GeCoA..53..197A}.
 \par\noindent
}\hss}}
\end{tabular}
\end{center}
\end{table}

\section{Discussion}
\label{section:discussion}

\subsection{1RXS J180556.1$-$343818}
\label{subsection:1RXSJ180556_discussion}
From the XMM-Newton observation, for the first time, we revealed the
broad-band ($0.3$--$7$~keV) X-ray spectrum of the 1RXS
J180556.1$-$343818.  The X-ray spectrum was well reproduced with an
absorbed optically-thin thermal plasma model with temperatures of $0.5$
and $1.7$~keV.  Although the best-fit absorption column density depends
on the model employed in the spectral fit, the $90$\% upper limit of
$N_{\rm H} = 9.6 \times 10^{20}$~cm$^{-2}$ obtained with the
multi-temperature plasma model was smaller than $1.0 \times
10^{21}$~cm$^{-2}$, a half of the total Galactic H\emissiontype{I}
column density in the line of sight to the source
\citep{1990ARA&A..28..215D}.  This smaller absorption suggests that 1RXS
J180556.1$-$343818 is a foreground source, not located in the Galactic
bulge.

Using the SIMBAD Astronomical
Database\footnote{http://simbad.u-strasbg.fr/simbad}, an optical
counterpart to 1RXS J180556.1$-$343818 was searched.  A rotationally
variable star, HD 321269, was found at $(RA, Dec)_{\rm J2000.0} =
(\timeform{18h05m56.454s}, \timeform{-34D38'29.89''})$ within the
pointing accuracy of XMM-Newton ($\sim \timeform{4''}$).  HD 321269 is
classified as a G9III object.  \citet{2000A&A...355L..27H} reported that
the $B$ and $V$-band magnitudes of this star were $10.67 \pm 0.05$~mag
and $9.884 \pm 0.059$~mag, respectively.  Assuming the absorption column
density of $N_{\rm H} = 3.6 \times 10^{20}$~cm$^{-2}$ to the source, the
color excess was calculated to be $E(B-V) = 0.062$~mag, using the ratio
of $N_{\rm H}$ to the color excess [$N_{\rm H}/E(B-V) = 5.8 \times
10^{21}$~atoms~cm$^{-2}$~mag$^{-1}$].  Thus, the intrinsic color index
of $B - V = 0.72$~mag was obtained.  Since the absolute $V$-band
magnitude for G giants was derived to be $\sim 0.8$~mag
\citep{1973asqu.book.....A}, we deduced the distance to the source of
$640$~pc.  Although the assumed parameters used in the above estimation
have relatively large uncertainties, the estimated distance reinforces
that the source belongs to the Galactic disk.

The X-ray luminosity in the $0.3$--$7$~keV is $6.9 \times 10^{31}
d^{2}_{640}$~erg~s$^{-1}$, where $d_{640}$ represents the source
distance in unit of $640$~pc.  This X-ray luminosity is comparable to
that of fast rotating G giants, e.g., FK Comae (G5III,
\cite{2002A&A...383..919G}) or $\beta$ Ceti (G9.5III,
\cite{1998A&A...330..139M}), which luminosities were found to be
$(3$--$10) \times 10^{31}$~erg~s$^{-1}$ in the $0.3$--$10$~keV band and
$(1.5$--$2.7) \times 10^{30}$~erg~s$^{-1}$ in the $0.2$--$4$~keV band,
respectively.

To elucidate plasma structures in stellar coronae, the distribution of
the emission measures (EMs) given as a function of temperature ($kT$)
has been extensively examined with high-resolution spectroscopic studies
using grating modules (e.g., \cite{2003A&A...404.1033A},
\cite{2004ApJ...617..531A}, \cite{2004A&A...413..643S},
\cite{2005ApJ...622..653T}, \cite{2013ApJ...768..135H}).  However, the
X-ray spectra with moderate energy resolution for FK Comae and $\beta$
Ceti are known to be well reproduced with a two-temperature plasma
model, similar to that for 1RXS J180556.1$-$343818.  The temperatures
were $0.7$ and $3.3$~keV for FK Comae during its low flux phase, and
$0.65$ and $1.0$~keV for $\beta$ Ceti.  The lower-temperature plasma
with $kT \sim 0.5$~keV of 1RXS J180556.1$-$343818 probably reflects a
solar-type active region, which corresponds to a peak at $\log T \sim
6.9$ of the EM distribution for G-type stars
\citep{2005A&A...432..671S}.

The temperature of the hot plasma ($kT = 1.7$~keV) of 1RXS
J180556.1$-$343818 was an intermediate value compared with those of
$\beta$ Ceti and FK Comae.  As a possible origin of these hot components
with $kT > 1$~keV, a continuous flaring activity is suggested (e.g.,
\cite{1997ApJ...480L.121G}).  It is still an open issue what are the
main physical conditions that determine X-ray flaring activities of
late-type giants.  However, we note that such flaring activities may be
somewhat related to the stellar rotation.  \citet{1981ApJ...248..279P}
indeed pointed out the correlation between the X-ray luminosity and the
projected rotational velocity ($v \sin i$) for late-type stars.
\citet{2006A&A...460..695T} reported that the rotational velocity of HD
321269 was $v \sin i = 33$~km~s$^{-1}$, while the rotational velocities
of FK Comae and $\beta$ Ceti were known to be $\sim 160$~km~s$^{-1}$ and
$\sim 3$~km~s$^{-1}$, respectively.

The elemental abundances of 1RXS J180556.1$-$343818 were found to be
depleted for Fe.  Regardless of the abundance tables and plasma
codes\footnote{Because of the different approach to Fe-L lines,
\texttt{apec} and \texttt{cvmekl} implemented in \texttt{XSPEC} cause a
slight difference in the spectra below $1$~keV. See also
http://www.atomdb.org/Issues/Fe\_Lambda\_ATOMDB\_SPEX.pdf} we used for
the spectral fit, the Fe abundance was significantly lower than those of
the other elements.  While O and Ne have first ionization potentials
(FIPs) higher than $10$~eV, the FIP of Fe is $7.9$~eV.  Thus, the
abundance pattern in the 1RXS J180556.1$-$343818 spectra may reflect
differences of the FIPs.  A similar abundance pattern was reported for
HR 1099 \citep{2001A&A...365L.324B}, in which low-FIP elements are
suppressed, compared with high-FIP ones.  Diffusive processes involved
with magnetic fields could operate elemental segregation in the corona
that leads to such an FIP effect.  In terms of the presence of a layer
with an excess of high-FIP elements, \citet{1993ApJ...408..373S}
explained the Ne and S enrichment in the flaring solar plasma.  Although
HR 1099 is an RS CVn binary, the rotational velocity of HR1099 was
obtained to be $v \sin i = 40$~km~s$^{-1}$ \citep{1999ApJS..121..547V},
comparable to that of HD 321269.  Thus, the rotational velocity of the
system may be a factor to influence the abundance pattern in the coronal
plasma.

\subsection{1RXS J173905.2$-$392615}
\label{subsection:1RXSJ173905_discussion}
\begin{figure}[htbp]
 \begin{center}
   \FigureFile(80mm,50mm){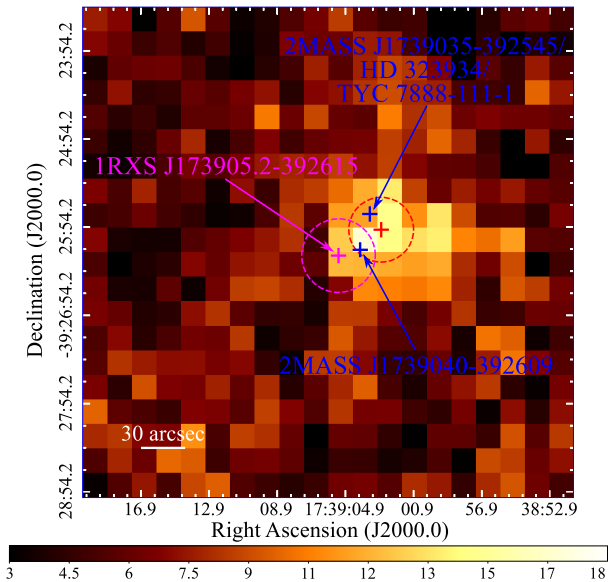}
 \end{center}
 \caption{Zoomed-up image of Suzaku J1739$-$3926.  The vignetting
correction, the binning of $16 \times 16$~pixels, and the smoothing with
$\sigma = \timeform{16.6''}$ were applied to the image.  The best-fit
position derived from the 2-dimensional PSF fitting is indicated with a
red plus.  The blue and magenta pluses represent the positions of 2MASS
J1739035$-$392545, 2MASS 1739040$-$392609 and 1RXS J173905.2$-$392615,
respectively.  The positional uncertainty including the absolute
accuracy of the Suzaku pointing is shown with a red dashed circle.
Meanwhile, a typical positional error of the RBSC sources was drawn with
a magenta dashed circle.  The color scales are in unit of photons per
$\timeform{16.6''} \times \timeform{16.6''}$ image pixel.}
\label{fig:1RXSJ173905_zoomup_image}
\end{figure}

In the Suzaku observation of 1RXS J173905.2$-$392615, we found faint
X-ray emission, designated as Suzaku J1739$-$3926, in the center of the
XIS field of view.  Judging from the radial profile of the surface
brightness, the X-ray emission probably originated from a single point
source.  The source position was given to be $(RA, Dec)_{\rm J2000.0} =
(\timeform{17h39m02.8s}, \timeform{-39D25'56''})$, indicated with a red
plus of figure~\ref{fig:1RXSJ173905_zoomup_image}.  However, the exact
position of the RBSC source (magenta plus in
figure~\ref{fig:1RXSJ173905_zoomup_image}) was located $\timeform{34''}$
away in the south-eastern direction from Suzaku J1739$-$3926.  The
positional accuracy of Suzaku J1739$-$3926 was estimated to be $\sim
\timeform{22''}$, taking into account the absolute pointing error in the
Suzaku observation, as is shown by a red dashed circle in
figure~\ref{fig:1RXSJ173905_zoomup_image}.  This circular region
overlapped with an error circle that represents a typical positional
accuracy of the RBSC sources ($\sim \timeform{25''}$ for the 90\%
confidence limit, \cite{1999A&A...349..389V}).  Hence, we concluded that
Suzaku J1739$-$3926 was consistent with 1RXS J173905.2$-$392615.

We note that $10$ near-infrared (NIR) sources, which are listed in the
2MASS Second Incremental Release Point Source Catalogue
\citep{2006AJ....131.1163S}, are positionally coincident with Suzaku
J1739$-$3926.  Since no further information was available for a unique
determination of the optical counterpart, hereafter, we only discuss the
possibilities of the counterpart for two relatively bright (with apparent
$J$-band magnitude of $\lesssim 11$~mag) sources: 2MASS
J1739035$-$392545 and 2MASS J1739040$-$392609.

2MASS J1739035$-$392545 is also identified with TYC 7888-111-1 and HD
323934 on the basis of the positional agreement.  The source location of
2MASS J1739035$-$392545 was $\timeform{13''}$ away from the best-fit
position of Suzaku J1739$-$3926 in the north-eastern direction (blue
plus in figure~\ref{fig:1RXSJ173905_zoomup_image}).  The $B$ and
$V$-band magnitudes of TYC 7888-111-1 were known to be $10.86 \pm
0.07$~mag and $10.95 \pm 0.12$~mag, respectively, and then the spectral
type of HD 323934/TYC 7888-111-1 was inferred to be A2.  However, an
unambiguous stellar classification is difficult;
\citet{2010PASP..122.1437P} recently derived the spectral type of F0II
to HD 323934/TYC 7888-111-1 by fitting a compilation of photometric
data.  If 2MASS J1739035$-$392545/HD 323934/TYC 7888-111-1 is indeed an
A-type star, we can estimate the luminosity distance as follows.
Assuming the upper limit of the absorption column density to be $N_{\rm
H} = 6.8 \times 10^{21}$~cm$^{-2}$, the interstellar extinction in the
$V$-band was estimated to be at most $A_{V} \sim 3.6$~mag [$A_{V} = 3.1
E(B - V) = N_{\rm H} / 1.9 \times 10^{21}~{\rm mag}$].  While the
apparent $V$-band magnitude was $m_{V} = 7.3$~mag, the absolute
magnitude of A2 main sequence stars is $M_{V} = 1.3$~mag.  Hence, the
luminosity distance was estimated to be $\sim 160$~pc.  We thus deduced
the X-ray luminosity in the $0.5$--$8$~keV band of $2.7 \times
10^{29}$~erg~s$^{-1}$.

It is well known that A or F-type stars, e.g., Altair
\citep{2009A&A...497..511R} and HR 8799 \citep{2010A&A...516A..38R}, are
faint X-ray sources with $\lesssim 10^{29}$~erg~s$^{-1}$ and seldom show
intense X-ray flares, as is explained by weak magnetic activity and
stellar winds.  Indeed, Altair and HR 8799 are stable in X-rays during
the decades.  However, assuming that the X-ray source is indeed 1RXS
J173905.2$-$392615 and that the ROSAT spectrum is reproduced with the
best-fit optically-thin thermal plasma model described in
section~\ref{subsection:1RXSJ173905}, the source shows the flux drop
from $6.6 \times 10^{-13}$ to $8.7 \times
10^{-14}$~erg~s$^{-1}$~cm$^{-2}$.  Hence, the X-ray emission may not be
attributed to 2MASS J1739035$-$392545.  A possible scenario to explain
this X-ray variability is that a hidden late-type companion of 2MASS
J1739035$-$392545 emits soft X-rays, similar to HD 161084
\citep{2008PASJ...60S..49M}, and that the companion was in an X-ray
flaring phase during the ROSAT observation.

2MASS J1739040$-$392609 is located $\timeform{20''}$ away from the
Suzaku J1739$-$3926 position in the south-eastern direction (another
blue plus in figure~\ref{fig:1RXSJ173905_zoomup_image}).  The $J$, $H$,
and $K$-band magnitudes of the source were $11.24 \pm 0.03$, $10.61 \pm
0.02$, and $10.38 \pm 0.03$~mag, respectively.  Since only the upper
limit of the absorption column density was obtained, we should take into
account a wide range of the stellar extinction ($A_{V} \lesssim
3.6$~mag).  Using the extinction ratios of $A_{J} / A_{K} = 2.50$,
$A_{H} / A_{K} = 1.55$, and $A_{V} / A_{K} \sim 8.8$
\citep{2005ApJ...619..931I}, the extinction in the respective bands were
$A_{J} \lesssim 1.0$, $A_{H} \lesssim 0.64$, and $A_{K} \lesssim
0.41$~mag.  Thus, the intrinsic color index of 2MASS J1739040$-$392609
was estimated to be $(J-H)_{0} = 0.2$--$0.6$~mag and $(H-K)_{0} =
0$--$0.2$~mag.  According to table 3 of \citet{2007AJ....134.2398C}, the
combination of the color indices corresponds to main sequence stars with
a spectral type of F--K.  Assuming an F5V star with an absolute
magnitude of $M_{J} \sim 2.6$~mag \citep{2007AJ....134.2398C}, the
luminosity distance of 2MASS J1739040$-$392609 is derived to be $\sim
330$~pc, and then we can infer the X-ray luminosity of Suzaku
J1739$-$3926 to be $1.1 \times 10^{30}$~erg~s$^{-1}$.  However, this
luminosity is slightly larger than those of canonical F-type stars.  We
also note that the $B$-band magnitude of 2MASS J1739040$-$392609 is
$15.2$~mag, much fainter than that of 2MASS J1739035$-$392545.  Thus,
2MASS J1739040$-$392609 may be a G or K late-type star, rather than an
F-type star.

Since the information obtained from the Suzaku observation was quite
limited, we cannot rule out the other possibility that Suzaku
J1739$-$3926 is not related to 1RXS J173905.2$-$392615 and that the RBSC
source became extremely faint below the sensitivity limit of the Suzaku
XIS.  The unique identification of the optical counterpart is an
indispensable approach to reveal the nature of 1RXS J173905.2$-$392615.
Therefore, the follow-up observation with Chandra would be encouraged to
determine the source position more accurately, using its unprecedented
high angular resolution of $< \timeform{0.5''}$.

\section{Summary}
\label{section:summary}
We conducted the XMM-Newton and Suzaku observations of the unidentified
X-ray sources towards the Galactic bulge, 1RXS J180556.1$-$343818 and
1RXS J173905.2$-$392615, in order to acquire the X-ray spectrum above
$2$~keV.  We deduced that these two sources are probably located in the
Galactic disk, not in the bulge.  The results of our image and spectral
analysis are summarized below.

\begin{enumerate}
 \item X-ray emission was detected at $(RA, Dec)_{\rm J2000.0} =
       (\timeform{18h05m56.3s}, \timeform{-34D38'29.3''})$, positionally
       coincident with a G-type giant, HD 321269, within the positional
       accuracy of XMM-Newton.  The X-ray spectrum of 1RXS
       J180556.1$-$343818 was well reproduced with an optically-thin
       thermal plasma model with two temperatures affected by the
       photoelectric absorption.  The best-fit temperatures were $0.5$
       and $1.7$~keV.  The X-ray luminosity in the $0.3$--$7$~keV band
       was estimated to be $6.9 \times 10^{31}
       d^{2}_{640}$~erg~s$^{-1}$, similar to that of rotating G giants.
       Compared with the O and Ne abundances, the Fe one was depleted,
       indicating the metal segregation.  The X-ray activity and the
       abundance pattern in the coronal plasma should reflect the
       stellar rotational activity.
 \item Dim X-ray emission that was positionally consistent with 1RXS
       J173905.2$-$392615 was detected.  Taking into account the
       contribution from the Galactic Ridge X-ray Emission, we made the
       X-ray spectrum in the $0.5$--$8$~keV band.  Due to the limited
       photon statistics, the meaningful constraints to the spectral
       parameters were not obtained.  The X-ray flux of $8.7 \times
       10^{-14}$~erg~s$^{-1}$~cm$^{-2}$ was an order of magnitude
       fainter than that during the ROSAT observation.
\end{enumerate}

\bigskip
We appreciate Dr. Kenji Hamaguchi and another anonymous referee for
their useful and careful comments to improve our manuscript.  We would
like to thank all the Suzaku team members for their support of the
observation and useful information on the XIS and HXD analyses.  This
work is also based on the observation obtained with XMM-Newton, an ESA
science mission with instruments and contributions directly funded by
ESA Member States and the USA (NASA).  Furthermore, this publication
makes use of data products from the Two Micron All Sky Survey, which is
a joint project of the University of Massachusetts and the Infrared
Processing and Analysis Center/California Institute of Technology,
funded by the National Aeronautics and Space Administration and the
National Science Foundation.

\end{document}